\documentclass{acm_proc_article-sp}
\usepackage[hidelinks]{hyperref}
\usepackage{comment}
\usepackage{array}
\usepackage{url}
\usepackage{listings}
\usepackage{algorithm}
\usepackage{algorithmic}
\usepackage{multirow}
\usepackage{subfig}
\usepackage{color}

\newcommand{\figref}[1]{Figure~\ref{#1}}
\newcommand{\secref}[1]{Section~\ref{#1}}

\newcommand{\tabref}[1]{Table~\ref{#1}}

\begin{document}

\title{A comparison of non-intrusive load monitoring methods for commercial and residential buildings}

\numberofauthors{5} %  in this sample file, there are a *total*
\author{
% 1st. author
\alignauthor
Nipun Batra\\
       \affaddr{Indraprastha Institue of Information Technology}\\
       \affaddr{Delhi, India}\\
       \email{nipunb@iiitd.ac.in}
% 2nd. author
\alignauthor
Oliver Parson\\
       \affaddr{University of Southampton}\\
       \affaddr{Southampton, UK}\\
       \email{osp@ecs.soton.ac.uk}
% 3rd. author
\alignauthor 
Mario Berges\\
       \affaddr{Carnegie Mellon University}\\
       \affaddr{Pittsburgh, Pennsylvania, USA}\\
       \email{mberges@andrew.cmu.edu}
\and  % use '\and' if you need 'another row' of author names
% 4th. author
\alignauthor Amarjeet Singh\\
        \affaddr{Indraprastha Institue of Information Technology}\\
       \affaddr{Delhi, India}\\
       \email{amarjeet@iiitd.ac.in}
% 5th. author
\alignauthor Alex Rogers\\
       \affaddr{University of Southampton}\\
       \affaddr{Southampton, UK}\\
       \email{acr@ecs.soton.ac.uk}
}

\maketitle
\begin{abstract}
Non intrusive load monitoring (NILM), or energy disaggregation, is the process of separating the total electricity consumption of a building as measured at single point into the building's constituent loads. Previous research in the field has mostly focused on residential buildings, and although the potential benefits of applying this technology to commercial buildings have been recognised since the field's conception, NILM in the commercial domain has been largely unexplored by the academic community. As a result of the heterogeneity of this section of the building stock (i.e., encompassing buildings as diverse as airports, malls and coffee shops), and hence the loads within them, many of the solutions developed for residential energy disaggregation do not apply directly. In this paper we highlight some insights for NILM in the commercial domain using data collected from a large smart meter deployment within an educational campus in Delhi, India, of which a subset of the data has been released for public use. We present an empirical characterisation of loads in commercial buildings, highlighting the differences in energy consumption and load characteristics between residential and commercial buildings. We assess the validity of the assumptions generally made by NILM solutions for residential buildings when applied to measurements from commercial facilities. Based on our observations, we discuss the required traits for a NILM system for commercial buildings, and run benchmark residential NILM algorithms on our data set to confirm our observations. To advance the research in commercial buildings energy disaggregation, we release a subset of our data set, called COMBED (commercial building energy data set).
\end{abstract}

%\category{I.5}{Pattern Recognition}{Applications}

%A category including the fourth, optional field follows...
%\category{I.2}{Artificial Intelligence}{Learning}[Parameter learning]  

%\keywords{energy disaggregation; non-intrusive load monitoring; smart meters}

\section{Introduction}
\label{introduction}
Energy disaggregation, also known as non-intrusive load monitoring (NILM), or non-intrusive appliance load monitoring (NIALM), can be broadly defined as a set of techniques used to obtain estimates of the electrical power consumption of individual appliances from measurements of voltage and/or current taken at a limited number of locations in the power distribution system of a building. The original ideas trace their roots back to the late 1980's, when George Hart at the Massachusetts Institute of Technology (MIT) in the US proposed using changes in real and reactive power consumption, as measured at the utility meter, to track the operation of individual appliances in a home automatically. Since then, a great number of commercial, academic and government efforts have developed and tested methods to solve this problem~\cite{armel_2013,zeifman_2011, zoha}.

The majority of the solutions developed to date have focused on residential buildings, despite the fact that the potential benefits of applying this technology to commercial and/or industrial buildings have been recognised from the field's foundation~\cite{norford1996}. Early assessments of the technology, detailed in reports by the Electric Power Research Institute in the US~\cite{epri_1997_1, epri_1997_2}, and later in a report prepared for the the California Energy Commission's Public Interest Energy Research (PIER) program~\cite{smith2003final}, concluded that there were issues with the technology's ability to deal with multi-state and variable-power devices, as well as multiple units of devices of the same make/model. These issues, although present to some extent in residential buildings, are much more prevalent in commercial facilities, which may explain why the technology development has focused primarily on the former.

Among the issues found by early assessments of the technology, the impact of complex load types on the performance of the algorithms is of particular importance. Two-state appliances (i.e., those that only operate in one of two states, typically {\em on} and {\em off}), are easier to track than those that have more complicated operations due to the number of states that they can be in and/or the possible transitions between those states. Variable loads, such as variable-speed drives, are one extreme form of a multi-state device.

Other types of loads that are more prevalent in commercial buildings can also complicate the disaggregation problem. For example, power factor correction devices, which introduce additional harmonics and noise into the building's power distribution system, fall into this category. This is especially true when high-frequency features are used to solve the disaggregation problem. Similarly, depending on the electrical point at which the measurements are taken, the integration of additional distributed energy resources 
%(DER)
such as micro-turbines, battery storage technologies (e.g., electric vehicles), and photovoltaic panels, are likely to impact the performance of load disaggregation algorithms.

There are, however, some positive aspects for deploying NILM solutions in this sector:

\begin{enumerate}
        \item Due to their complexity and magnitude, commercial buildings are typically managed by an experienced facility management services crew. These are likely to be the primary customers and users of NILM technologies, and given that this solution directly impacts their daily work activities, they are more likely to be motivated to embrace the technology.
        \item Commercial buildings are more likely to have a detailed asset management system, which can be used to assist in creating the database of appliance signatures for the NILM system.
        \item Commercial facilities are more likely to have existing building automation systems, 
        %(BAS)
        that are used to monitor and control specific building systems. These measurement and control points can provide very useful information that can assist the disaggregation engine.
        \item There are significantly more opportunities for performing automated fault detection and diagnosis and system-specific efficiency analyses.
        \item A higher return on investment 
        %(ROI) 
        on each measurement point is more likely, given the value of the assets present in commercial facilities, and the amount of energy consumed by the devices.
        \item Some commercial facilities tend to have more regular, periodic and predictable operation given that they are based on fixed work schedules.
        \item Due to the nature of the services being provided by these facilities, automated control of end-use loads is more likely to be accepted and sought after by the building constituents (owners, managers, occupants). This, in turn, increases the value proposition for the hardware installations.
		\item Regular energy audits for several commercial units (e.g. cement industry in India) are often mandated for compliance purposes. Detailed analysis from NILM can further help in simpler execution of cumbersome energy audits that currently require multiple measurements across the facility.
\end{enumerate}

In this paper, we explore these challenges and opportunities in more detail, by making use of a year-long data set collected at an academic campus with dense sub-metering. While there exists a vast diversity of commercial buildings, such as airports, IT offices and malls, we shall focus on IT offices, due to the lack of publicly available data for other building types. Many of the buildings from our data set can be considered to be similar to IT offices where HVAC and IT loads are the primary energy consumers. We use our data set to highlight the key differences in energy consumption between residential and commercial buildings, specifically from a NILM perspective. Specifically, the primary contributions of our work are:

\begin{enumerate}
\item We describe a data set collected from an extensive deployment of smart meters across an educational campus in India. A subset of our data set, for one building, constituting data at different levels (buildings, floors and major loads), is released with this paper.
\item We assess the validity of the assumptions generally made by NILM solutions for residential buildings, when applied to commercial facilities. %This validation is performed using a data set collected from an extensive deployment of smart meters across an educational campus in India. 
\item We perform an empirical characterisation of the major loads in our data set, providing insights into the traits for a NILM system for commercial buildings. 
\item We apply benchmark residential NILM algorithms to our data set to further validate the developed insights. This is the first time that these well known NILM algorithms have been applied to a commercial setting. 
%\item A subset of our data set, for one building, constituting data at different hierarchies - building, floor and major loads, is released with this paper. We believe this is the first such data set from a commercial building and will help advance the state of NILM for commercial buildings. 
\end{enumerate}

The remainder of this paper is structured as follows. In \secref{related_work} we discuss relevant energy disaggregation work from both the commercial and the residential domains. In \secref{assumptions} we introduce the set of assumptions that are typically made in NILM solutions for residential settings. We then describe our deployment in a university campus in Delhi, India in \secref{deployment}. Data from our deployment, together with two publicly released data sets for residential NILM are used to validate the applicability of residential NILM methods for commercial setting in \secref{comparison}. As part of this validation, we further develop insights into useful traits for commercial NILM methods. Thereafter, in \secref{evaluation}, we apply benchmark residential NILM algorithms to our data and validate the developed insights. Finally, conclusions and future work are discussed in \secref{conclusion}.

\section{Related work}\label{related_work}

The most significant work which has applied energy disaggregation to commercial buildings was completed at MIT during 2001-2003. Norford, Xing and Luo presented an approach based on the generalised likelihood ratio and applied it to disaggregate a set of chillers, fans and pumps in a single story academic test building~\cite{norford_2001}. The authors collected aggregate and sub-metered data at 24~Hz, and identified 1 second as the minimum sample rate required to identify most appliance switch events. However, the approach required sub-metered appliance data to first be collected for training purposes, and furthermore the authors describe how the proposed approach was unable to detect appliances containing variable speed drives, such as fans, due the start up transients which lasted multiple minutes.

Norford and Lee later applied an extension of the previous approach to one floor of an office building~\cite{norford_2001a}. The authors collected aggregate current and voltage data at 120~Hz, from which the proposed approach was able to distinguish between the different HVAC loads that were in use. However, the approach required a manual training phase in which each load was operated multiple times in isolation of the operation of other appliances. In addition, the authors also noted the difficulty of identifying slowly varying loads or multi-step change loads.

Lee, Norford and Leeb then went on to further extend this work to use both step changes in the aggregate load and longer transients to disaggregate the various loads within a building~\cite{lee_2003}. The authors collected aggregate current and voltage data at 16~kHz from two commercial buildings: one floor of an office building and the laundry room of a dormitory. The proposed approach made extensive use of high frequency data, from which higher order harmonics could be extracted from the current signal, allowing loads to be disaggregated to a higher level of accuracy. However, the authors again highlight the need for a manual training phase required to first learn the behaviour of each load before their approach was able to disaggregate the loads. Lee et al. have also developed Gaussian random process models for loads with variable speed drives by exploiting the correlations between mean harmonic powers~\cite{lee2005estimation}. Recently, Shao et al.~\cite{shao} used temporal motif mining techniques for load disaggregation in both residential and commercial settings.Their approach exploits the fact that some loads in commercial buildings are run on regular schedules. While their approach is unsupervised, it requires one to specify the number of appliances which may be present in a home or the commercial building.

In summary, load disaggregation research for commercial buildings has increasingly focused on high-frequency (kHz) methods which require a manual training phase for each building due to the lack of generality of individual load features. In contrast, recent relevant work in the residential domain has started to apply low-frequency (< 1~Hz) methods to individual households which do not require a manual training phase. Such developments have been fuelled by the release of a range of residential data sets designed specifically for energy disaggregation, including REDD~\cite{redd}, BLUED~\cite{blued}, Smart*~\cite{smart}, AMPds~\cite{ampds}, iAWE~\cite{iawe}, ECO~\cite{eco}, GREEND~\cite{greend}, Sust-data~\cite{sustdata} and UK-DALE~\cite{UK-DALE}. To the best of our knowledge, only two commercial data sets have been released to date. While Enernoc\footnote{\url{http://open.enernoc.com/data/}} collected data across different industries, the data set contains only aggregate level readings and is sampled at 5 minute resolution. More recently, the BERDS~\cite{maasoumyberds} data set was released, which provides the power consumption of a selected number of loads at 15 minute resolution. Neither of these data sets contain 1~Hz data from the majority of appliances in the monitored buildings, and therefore have limited the progress of NILM research in the commercial domain.
%which are two traits present in the residential data sets discussed above. 

\section{Assumptions in residential NILM methods}\label{assumptions}
Having discussed relevant literature for NILM in commercial settings, we now introduce four key assumptions that are often made by residential disaggregation methods. %Based on this comparison we discuss the traits of NILM systems in the commercial settings.

\begin{figure*}
  \centering
  \includegraphics[scale=0.55]{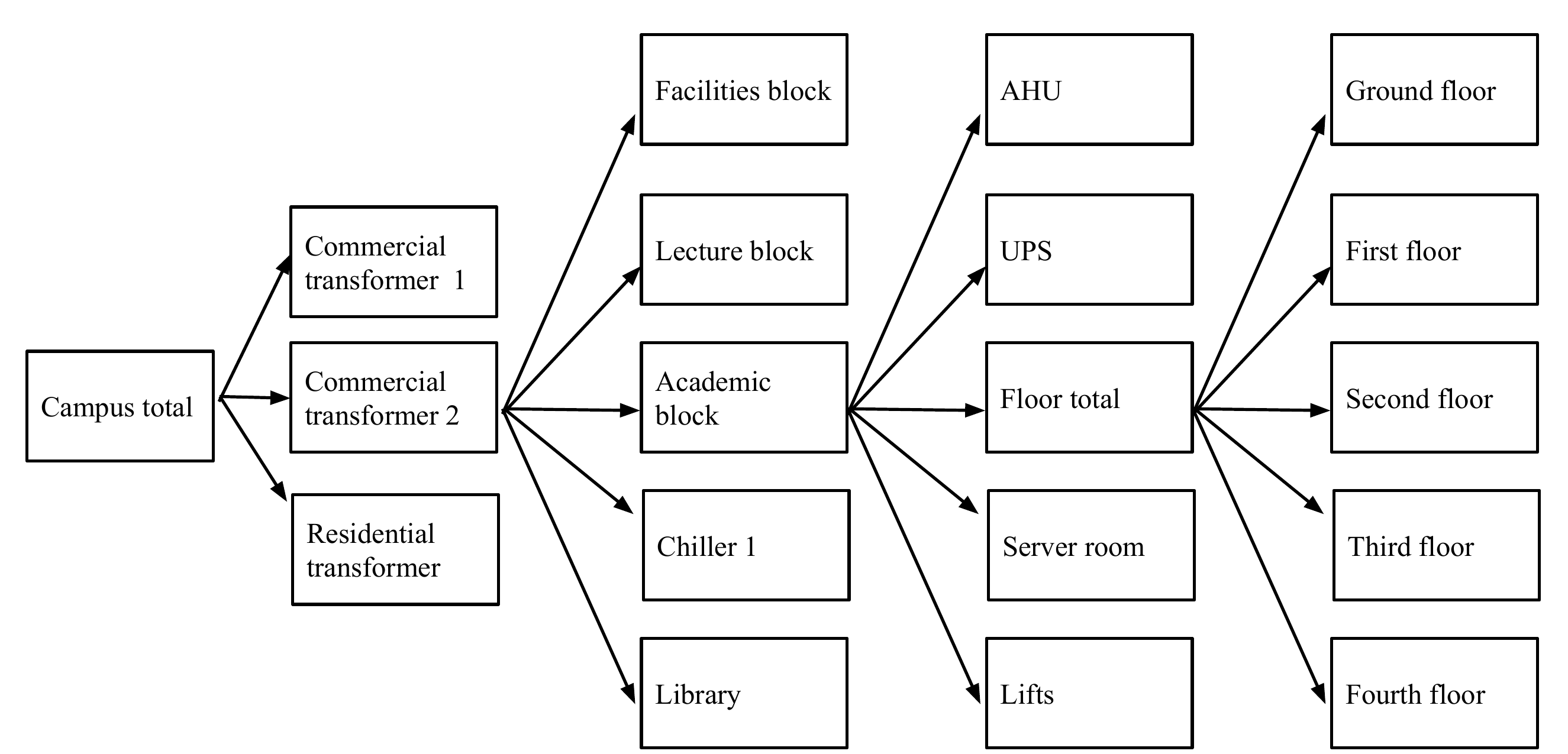} 
  \caption{An illustration of hierarchical electricity metering undertaken at IIIT Delhi, India.}
  \label{fig:metering} 
\end{figure*}

\textbf{One-at-a-time assumption:} Many energy disaggregation methods applied within the residential domain aim to detect appliance switch events from aggregate data. However, it is possible that the switch events of two or more loads might overlap (i.e., multiple appliances change state between two sequential samples of the building's aggregate load). Furthermore, the likelihood of overlapping appliance switch events increases as the sampling rate decreases. %This creates a computationally complex problem for disaggregation systems, since the overlap of any combination of appliance switch events could theoretically appear in the building's aggregate load. 
Accordingly, many residential NILM approaches make the one-at-a-time assumption, i.e.\ only one appliance will change state between two consecutive samples of the aggregate load~\cite{kolter_2012, liang_2010a}. Although such overlaps rarely occur given a sufficiently high sampling rate, it is important that disaggregation systems operating under this assumption may still generate only minor errors as a result of overlapping events rather than failing to accurately disaggregate any appliances.

\textbf{Steady-state loads:} A common feature of many residential loads is that such appliances draw a roughly constant power while remaining in the same state (often referred to as a steady-state). This feature allows an aggregate load to be conveniently divided into periods during which each appliance remains in a single state, separated by candidates where an appliance state change could occur~\cite{johnson_2013}. Such approaches cannot be applied to disaggregate appliances with a continuously variable load, such as plasma televisions, and furthermore often fail to disaggregate any appliances in households containing such variable loads. Such loads are relatively rare in residential buildings. However, variable speed pumps and fans are often used in HVAC systems for commercial buildings instead of fixed speed motors, as a result of their reduced energy consumption. These variable speed pumps exhibit prolonged transients which would cause steady-state detectors, that assume short duration transients, to fail~\cite{norford1996}.

\textbf{Temporal dependencies:} Residential appliance usage often follows a repeating daily and weekly schedule, as a result of the lifestyle of the household's occupants. Such repetitive behaviour has been exploited by disaggregation approaches, and has been shown to increase the accuracy of energy disaggregation algorithms~\cite{kim_2011,kolter_2010}. While such daily and weekly periodicity is common in residential buildings, the individual schedules for each household often need to be learned separately due to the high variance between households. In contrast, commercial building types are more likely to have similar daily or weekly schedules (e.g., a campus of buildings which share the same opening hours or manufacturing units often operating under the same shift patterns). Previous work~\cite{shao} also suggests exploiting fixed appliance schedules in commercial buildings. As a result, separately learning periodicity for each commercial building is likely not required. 

\textbf{Feed correlations:} In addition to temporal dependencies, strong correlations between appliance usages have also been exploited in residential disaggregation. For example, the usage of appliances such as televisions and set top boxes are often highly correlated, and similarly so are the usage of computers and monitors. It has been shown that such information can be exploited by a disaggregation algorithm to disaggregate appliances even when the step changes have not been detected~\cite{kim_2011}. However, perfect correlations between appliances can result in simultaneous start up or shut down events being produced, as might occur in commercial buildings, therefore contributing systematic violations of the one-at-a-time assumption.

We later evaluate these assumptions for commercial buildings in \secref{comparison} using two publicly available residential NILM data sets and the data set from our deployment which we describe next.

\section{Deployment}\label{deployment}
IIIT Delhi is a modern educational campus established in 2008 and spreads over 25 acres. The campus consists of 8 buildings: an academic block, lecture block, library, facilities building, mess building, faculty housing and male and female dormitories. We leveraged the construction phase of our campus to extensively deploy smart meters across the campus. More than 200 Modbus enabled Schneider Electric EM6400 and EM6436 smart electricity meters were deployed across different buildings and sub-systems spread throughout the campus.

Based on our previous successful deployments of electricity and ambient monitoring equipment~\cite{issnip,iawe}, we chose Raspberry Pi as our controller to collect data from Modbus enabled smart meters. %Since we powered the Raspberry Pis via the mains electricity supply as opposed to a battery based solution, we avoided the usual maintenance associated with battery based solutions. 
We chose sMAP~\cite{smap} as our data archiving and metadata description backend, as it is a mature platform previously proposed in the research community and has been used by other research groups\footnote{For example: \url{http://new.openbms.org/plot/}} for similar deployments.

Our deployment is also inspired by previous campus testbeds established by the community~\cite{dashboard_1, dashboard_2}. With the aim of creating a testbed for coming years, we decided to extensively instrument our campus at various metering levels. \figref{fig:metering} shows a simplified view of our smart meter deployment. We instrumented the 3 transformers (Commercial transformers 1 and 2, and Residential transformer in \figref{fig:metering}) which are the entry points from the utility. The residential transformer supplies the faculty housing and the dormitores, while the other two transformers cater to the remaining five buildings and all central loads (such as HVAC chillers, HVAC pumps and UPS). For each building, we separately monitor total power consumption, high energy consumption loads (e.g. AHU, elevators and UPS) and total consumption for each floor. In addition, certain central loads that cater to the whole campus, such as the HVAC chiller or HVAC pumps, which have a high energy consumption were also separately monitored. Many of the non-residential buildings (e.g., academic block and library) can be considered to be similar to IT offices where HVAC and IT  loads are the main energy consumers. %Within each of these buildings, we monitored all the major HVAC loads (such as the air handler unit (AHU), air washer, etc.). Each of these buildings also had dedicated UPS for critical IT equipment and dedicated supplies for lighting, both of which were monitored. Lastly, we also monitored the power consumption of individual floors in each of these buildings. Additionally, in some of the buildings, we went a level further to measure wing level power consumption across some of the floors.
The campus deployment is summarised in \tabref{tab:deployment}. We have released one month of data from the academic block constituting data at different hierarchies (buildings, floors and major loads) for public use. More details about our data set, called COMBED (commercial building energy data set) can be found on its project page.\footnote{\url{http://combed.github.io}}

\begin{table}
\begin{center}
\begin{tabular}{c c}
\hline
\textbf{Data set attribute}&\textbf{Value}\\
\hline
\# Smart meters & 200\\
\# Electrical parameters measured & 8\\
\# Buildings instrumented & 6\\
Sampling interval & 30 s\\
Deployment period & Aug 2013-present\\
\hline
\end{tabular}
\end{center}
\caption{Summary of our campus deployment}
  \label{tab:deployment}
\end{table}

\begin{figure*}[!htb]
  \centering
  \includegraphics[scale=1]{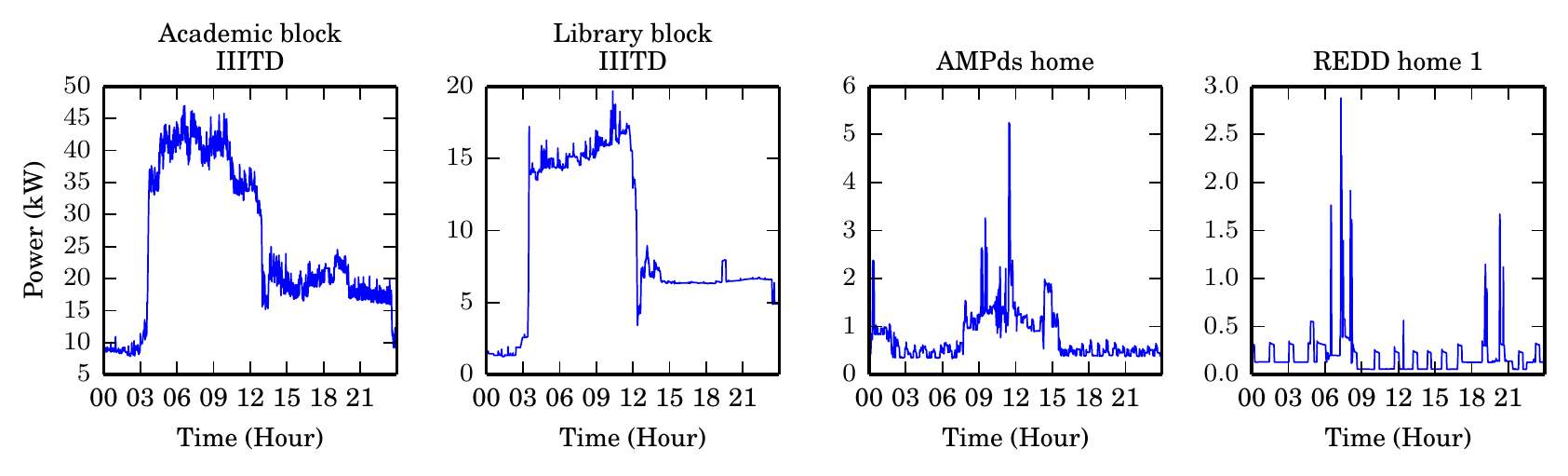} 
  \caption{Comparison of power consumption in residential and commercial buildings.}
  \label{fig:total_power} 
\end{figure*}

\begin{figure*}[!htb]
  \centering
  \includegraphics[scale=1.15]{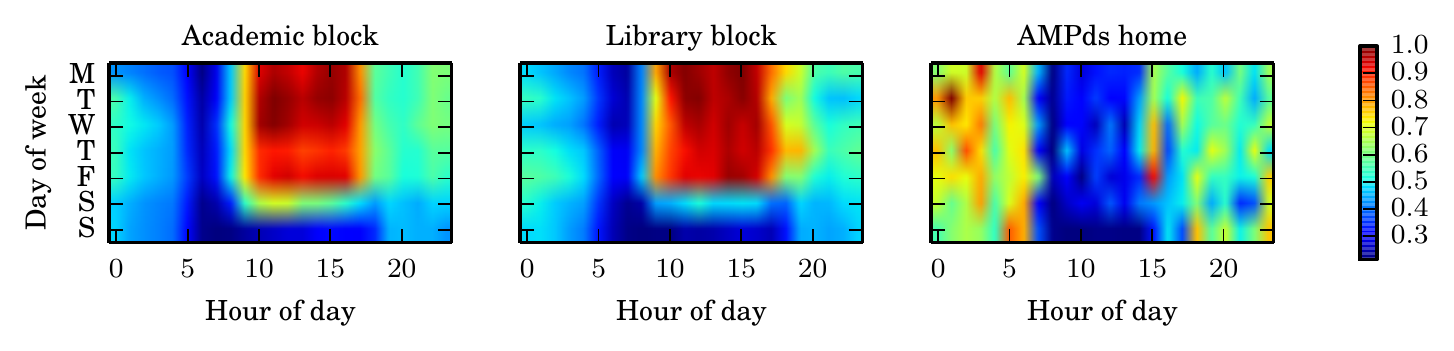} 
  \caption{Hourwise energy consumption of academic and library blocks from our deployment compared to the home from AMPds data set. The energy consumption has been normalised with respect to the maximum energy consumption for each building to facilitate qualitative comparison. Energy consumption increases from blue to red.}
  \label{fig:heatmap} 
\end{figure*}

\section{Comparison of residential and commercial buildings}\label{comparison}
%Having described our campus deployment, we now compare the NILM related problems in these settings with the more studied residential settings. 
%\redcolor{One thing that is currently missing in this section is that each of the assumption we discuss earlier is not being referenced. Please see to it that you discuss each of the assumption earlier discussed to have a better correlation.}

We now evaluate the common residential NILM assumptions introduced in \secref{assumptions} in the context of our commercial data set described in \secref{deployment}.
Wherever possible, we make use of previously collected residential data sets while drawing comparisons. We believe that the release of such data sets has been the catalyst behind the recent expansion of the field, and therefore data sets in commercial settings might have a similar effect. As discussed previously in \secref{introduction}, \ref{related_work} and \ref{deployment}, many of the buildings from our campus closely resemble common office buildings and therefore many of our findings can be generalised to other IT office buildings. We begin our analyses by comparing the total energy consumption of residential and commercial buildings.

\subsection{Total energy consumption and temporal dependencies} 
\figref{fig:total_power} compares the total power demand observed on a particular day across the academic and library blocks (from our deployment) with homes from the AMPds and REDD residential data sets. As expected, commercial buildings consume considerably more power than their residential counterparts throughout the day due to the presence of a greater number of loads and high power consumption systems (such as HVAC chillers and pumps). In fact, the total energy consumption on these specific days is 612 kWh and 217 kWh for the academic block and library block, and 19 kWh and 5 kWh for the AMPds and REDD homes respectively. 
%As expected, the consumption varies by an order of magnitude, 
This confirms our initial intuition, which may indicate a greater potential for energy savings in the commercial settings, through the identification of inefficient appliances or usage schedules.

Previously, in \secref{related_work}, we discussed how NILM approaches designed for residential buildings have used temporal dependencies in the data to enhance disaggregation performance. We now contrast the temporal dependencies in the residential and commercial settings. \figref{fig:heatmap} shows the hour-wise normalised energy consumption of two buildings from our deployment (academic block and library block) compared to the home from the AMPds data set. We observe that the two buildings from our deployment consume significantly greater amounts of energy during office hours (9~AM to 5~PM) on weekdays in comparison to other time periods. In contrast, the energy consumption of the home from the AMPds data set shows some consistent variation between hours of day but 
%is much more consistent and 
does not show any significant variation between weekdays and weekends. While the residential data shows some temporal dependencies (such as lower energy consumption when people leave for office), the effect (as observed in \figref{fig:heatmap}) is much more profound in the commercial settings.

\textbf{Insights:} Previous studies have proposed using separate NILM models for weekdays and weekends. While residential buildings may show some difference in the energy consumption between different times of day, the difference is much more prevalent in the case of commercial buildings. Based on this observation, we believe that the day of the week is an important feature to consider when performing disaggregation. In our particular data set, we believe NILM performance accuracy is likely to improve if separate models for weekdays and weekends are considered, due to the fact that our campus runs classes only on weekdays.

\begin{figure*}[!htb]
  \centering
  \includegraphics[scale=0.9]{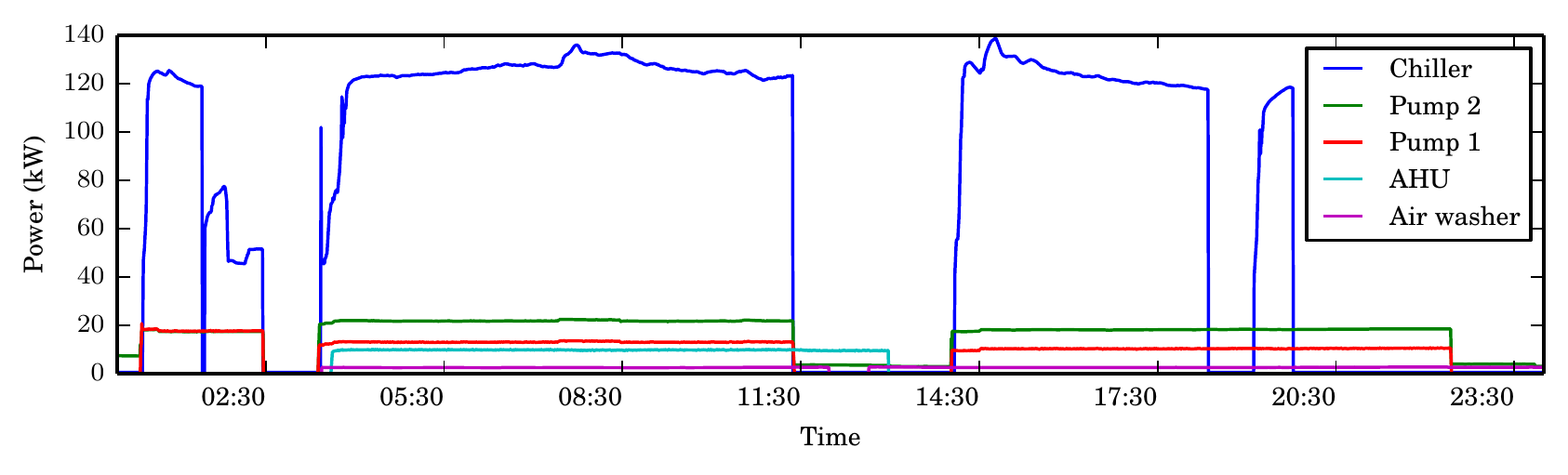} 

  \caption{Multiple HVAC systems operating at the same time invalidating the ``one-at-a-time" assumption made by residential NILM methods.}
  
  \label{fig:hvac} 
\end{figure*}

\subsection{Number of events} 
\label{events}
Commercial buildings usually contain many more loads in comparison to their residential counterparts. Thus, it is expected that there would be more frequent appliance activity and correspondingly more appliance switch events in commercial settings. This higher activity frequency is expected to generate a higher number of simultaneous appliance switch events as a result of random collisions. Based on this intuition we now evaluate the validity of the one-at-a-time assumption discussed in \secref{related_work}. 
%\redcolor{I dont understand how you are evaluating one-at-a-time assumption unless you state how many of these events indeed constitute overlapping appliances. Just showing high number of events does not say anything about overlapping events.} 
We define that an event has occurred when the power draw between successive samples varies by a value greater than some threshold. \tabref{tab:events} compares the number of events in a day for several buildings from our deployment to all 6 homes from the REDD data set, where a threshold of 100~W was chosen to define an event. We observe that the number of events in a commercial building can be two orders of magnitude greater than that of the homes in the REDD data set. Furthermore, as we move down the electrical metering hierarchy (from the transformer to the floor level), we see that groups of events become divided, signifying that the disaggregation problem is simplified by deploying additional meters. However, even at the floor level, the number of events in the commercial building are several times more than those observed in the residential settings.

If we increase the threshold which defines the events, the number of events reduces as shown in \tabref{tab:events_threshold}. We believe that while extending existing event-based NILM approaches to commercial buildings, the threshold parameter may need to be carefully fine-tuned. By increasing the threshold, we reduce the number of events caused by continuously varying loads such as AHUs (discussed in \secref{variability}) and also avoid small power events (like light switching) that may not be of much significance in commercial settings.

\textbf{Insights:} 
The one-at-a-time assumption often made in the context of residential NILM approaches is not expected to be valid in the commercial context due to the increased frequency of events. However, by adding more metering (transformer to building to floor), the number of simultaneous switch events is likely to decrease as the events get partitioned across different entities (e.g., floors, buildings, etc.). Furthermore, due to the presence of different categories of loads (larger power draw, varying power draw) in commercial buildings, the threshold for defining events might be significantly greater than their residential counterparts. We revisit the utility of this additional sub-metering in \secref{evaluation}.

\begin{table}[!t]
\begin{center}

\begin{tabular}{lrr}
\hline
\textbf{Building/Entity} &   \textbf{median} &  \textbf{max} \\ \hline
\textbf{Transformers}   &       &         \\
Transformer 1           &  2444.0 &  2596 \\
Transformer 2           &  2520.0 &  2636 \\
Transformer residential &  2678.0 &  2778 \\ \hline
\textbf{Campus buildings}&        & \\
Academic block          &  2050.5 &  2270 \\
Library block           &   753.0 &  1017 \\ \hline
\textbf{Floors in academic block} \\
First floor             &   648.5 &  1274 \\
Ground floor            &  1263.0 &  1450 \\
Second floor            &   184.5 &   366 \\ \hline
\textbf{Residential from REDD} \\ %\hline
Home 1                 &    67.0 &   259 \\
Home 2                 &    91.0 &   126 \\
Home 3                 &   87.0 & 285 \\
Home 4                 &    29.0 & 102\\
Home 5                 &    18.5 & 62\\
Home 6                 &    27.0  & 101\\
\hline
\end{tabular}
\end{center}
\caption{Comparison of number of events in a day in commercial and residential buildings. For a valid comparison we downsampled the REDD data set from 1 second resolution to 30 second resolution to match the sampling period from our deployment. }
\label{tab:events}
  
\end{table}

\tabcolsep=0.06cm
\begin{table}[!t]
\begin{center}
\begin{tabular}{lrrrrrr}
\hline

%\bf{Threshold~(W)} &  \bf{100}  &  \bf{200}  &  \bf{500}  &  \bf{1000} &  \bf{2000} & \bf{5000} \\ \hline

\multicolumn{1}{c}{\bf{Building}} &  \multicolumn{6}{c}{\bf{Threshold (W)}} \\
 &  100  &  200  &  500  &  1000 &  2000 & 5000 \\ \hline

Academic block &  2050 &  1808 &  1433 &  1083 &   754 &   222 \\
Library block  &   753 &   596 &   377 &   223 &   134 &    26 \\ \hline
\end{tabular}

\end{center}
\caption{Median of the number of events observed in the academic and library block for varying power thresholds. }
\label{tab:events_threshold}
  
\end{table}

\subsection{Correlated data streams} 
Commercial buildings often house several subsystems which operate together. \figref{fig:hvac} shows different HVAC components which usually turn on and turn off at the same time. This operation is likely triggered by a fixed schedule and is therefore independent of human activity. For instance, IIITD follows a policy of fixed HVAC system running times across multiple buildings. In contrast, in residential buildings, appliances operating together are often due to human activity. For example, the television and fan may be used together during the dinner time. As a result, the use of such feed correlations are likely to contribute greater increases to the disaggregation accuracy for commercial buildings than residential buildings, as discussed in \secref{related_work}.

%However, multiple appliances/systems turning on/off together also contribute systematic violations of the one-at-a-time assumption, as discussed in \secref{related_work}.

\begin{figure}[!t]
  \centering
  \includegraphics[scale=1]{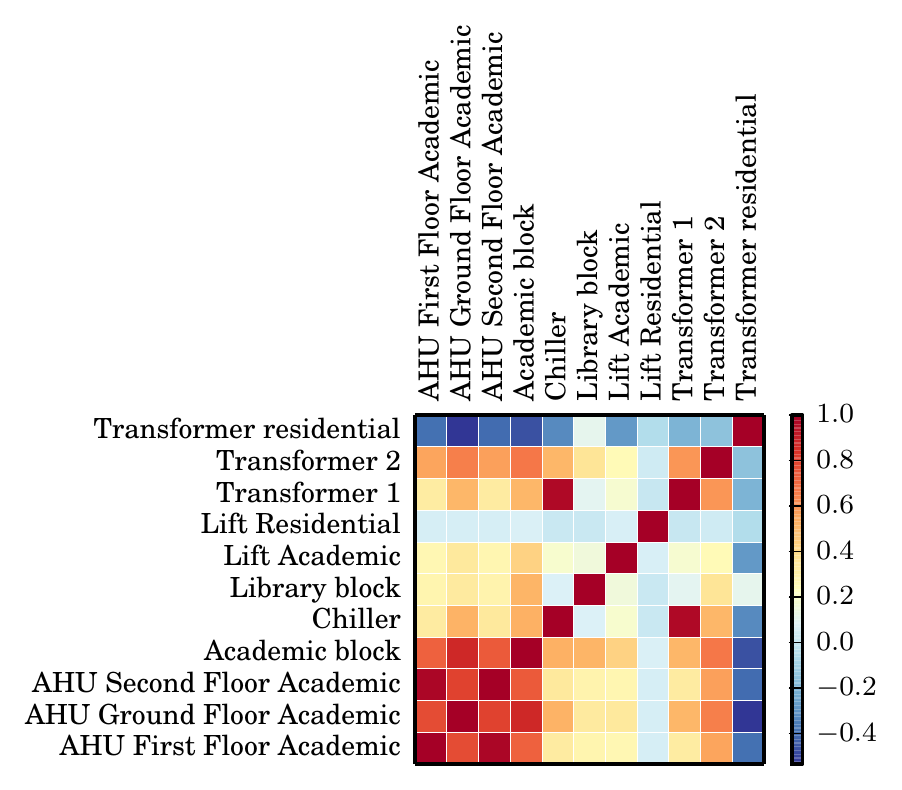} 
  \caption{Correlation heatmap between different subsystems from our deployment. +1 indicates total positive correlation and -1 indicates a total negative correlation.}
  \label{fig:corr} 
  
\end{figure}

\figref{fig:corr} shows the correlation between the power consumption of several streams from our deployment. The high correlation amongst the different HVAC systems (AHUs on different floors of the academic block and the chiller) clearly shows that these systems often operate together. In contrast, there is little correlation between the power draw of HVAC systems and the systems catering to the residential block (residential elevators and transformer). Recent research~\cite{fontugne2013strip} has also looked into sophisticated variants of stream correlation to identify abnormal energy consumption in buildings. We noticed the high correlation between transformer 1 and the chiller, which is a result of the chiller consuming the majority of energy supplied by this transformer. Such correlations could be used to automatically construct the metering hierarchy, by determining loads being monitored at multiple metering levels.
%suggesting that the chiller is the major load on this transformer. 
%We later confirmed that the chiller is the largest load in the campus and dominates the power consumption on the transformer from which it draws power.

\textbf{Insights:}
Although the strong correlations between feeds in commerical buildings can be used to increase the accuracy of disaggregation algorithms, such behaviour also consistently violates the one-at-a-time assumption. As a result, NILM methods which rely on edge matching, which would consider a rising or falling edge in the aggregate power signal to be caused by a single appliance, may need to be modified to account for this. However as previously suggested by Kim et al.~\cite{kim_2011} in the context of residential settings, the correlation among different systems can also be used to improve disaggregation performance. For example, if multiple loads in different buildings often change state simultaneously, it is highly likely to be caused by different HVAC components (AHUs, chillers, pumps) in these buildings which are often operated synchronously by the facilities.
% shouldn't this be shared HVAC commponents?

\begin{figure}[!t]
  \centering
  \includegraphics[scale=1]{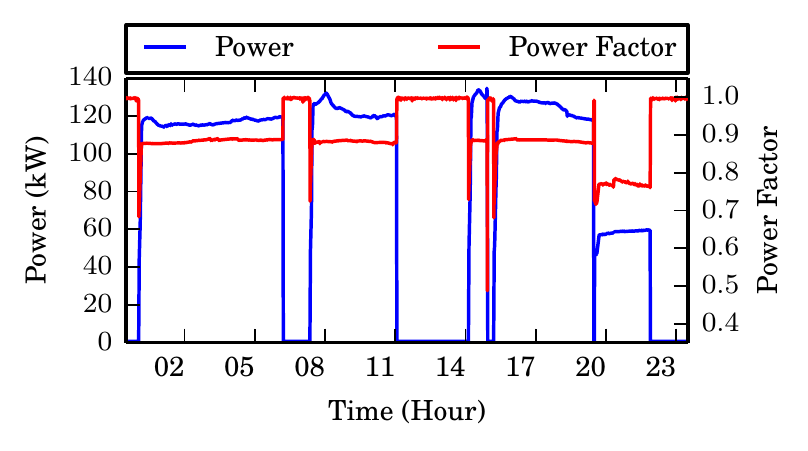} 
  \caption{Relationship between the power draw and power factor of our campus chiller. The power factor drops significantly (from 1 to 0.9) when the chiller is on.}
  
  \label{fig:pf} 
\end{figure}

Another insight stems from the fact that commercial buildings often pay additional charges for their peak power demand usage~\cite{compre}. The different HVAC systems starting up simultaneously all have a cumulative effect on the peak power demand. Thus, the company might incur heavy demand charges, therefore motivating the building manager to schedule the loads sequentially to lower the peak to reduce these costs. Furthermore, HVAC systems run on motors that are inductive in nature, and hence reduce the overall power factor, as shown by \figref{fig:pf}. Commercial buildings often use additional capacitor banks to compensate for the power factor reduction caused by inductive loads. Quick switching of large inductive loads results in quick switching of multiple capacitors in the capacitor bank that further reduces their lifetime and increase their chances of failure. Thus, distributing the operations of motorised loads also helps to improve the lifetime of the capacitive bank.

\subsection{Variability in power draw}\label{variability}

\begin{figure*}[!htb]
  \centering
  \includegraphics[scale=1]{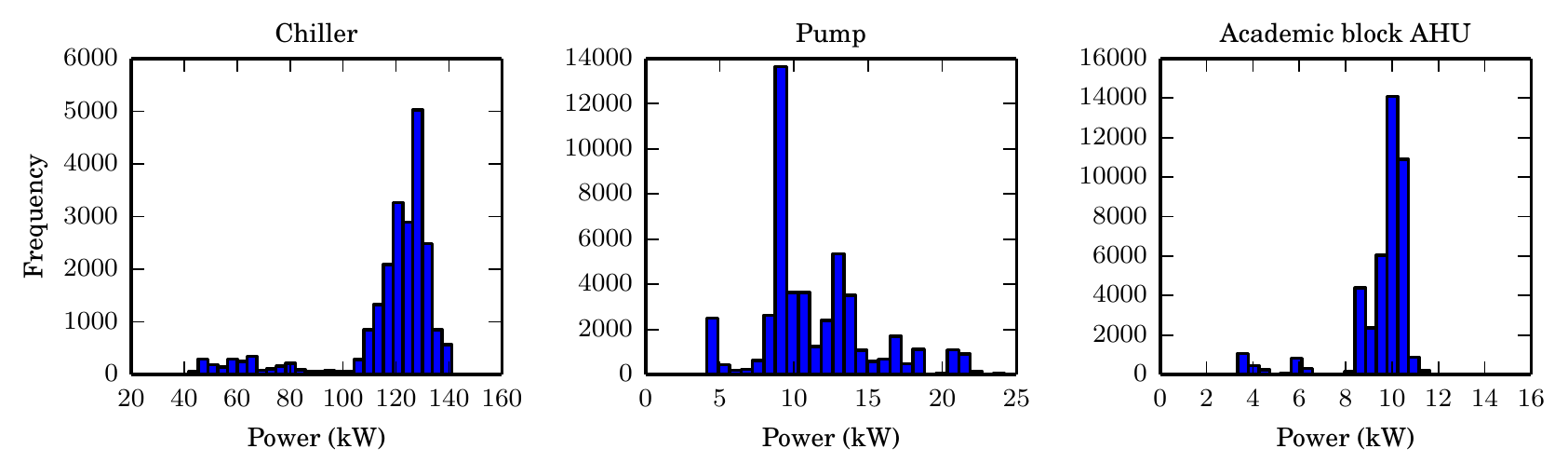} 
  \caption{HVAC components may employ Variable Frequency Drives (VFD) for optimising energy and performance that also result in variable power consumption}
  \label{fig:variable} 
\end{figure*}

Commercial buildings often employ variable frequency drives (VFDs) for optimising the energy and operation of HVAC systems~\cite{vfd_1, vfd_2}.  These VFDs allow changes in the fan speed in accordance with the desired load to achieve the set-point temperature for different HVAC systems, and hence optimise their energy consumption. \figref{fig:variable} shows a histogram of the instantaneous power draw of different HVAC systems from our campus deployment over several months of operation. It can be seen that these loads draw a highly variable and continuously varying power demand, and as such are expected to be difficult to disaggregate. As discussed previously in \secref{related_work}, approaches for residential buildings have often modelled appliances using a set of steady states, which would be a poor representation for such variable loads. The appliances which don't conform to steady state behavior in residential buildings include electronic equipment such as laptops and plasma televisions, few of which have a significant energy consumption. In contrast, the VFD-based systems used in commercial buildings have a significant contribution to the overall energy consumption and thus their accurate disaggregation is important.

\begin{figure*}[!htb]
  \centering
  \includegraphics[scale=1]{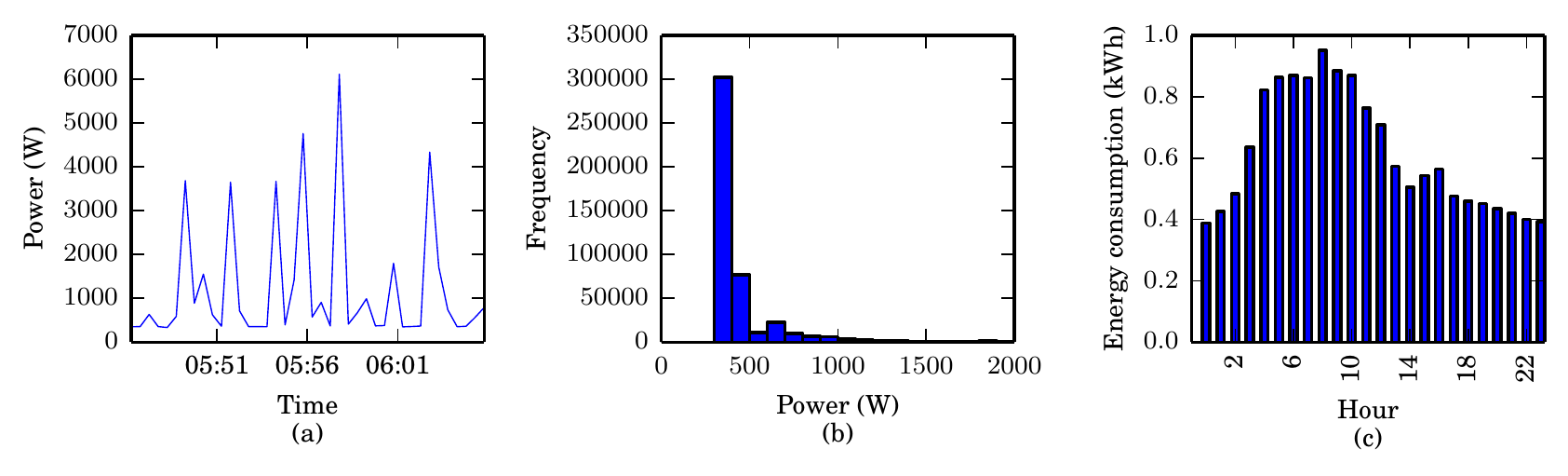} 
  %\caption{Elevators exhibit short term transient power consumption (a). Elevators mostly consume base load power (b) which is significant in comparison to their peak consumption typically seen during working hours (c). \redcolor{Part-c does not show peak but average. Either you need to replace the figure or reword what you are saying.}}
  \caption{(a) Short term transient power of two elevator loads. (b) Histogram of elevator power demand with a peak corresponding to base load. (c) Daily variation of elevator energy consumption indicating high contribution of base load.}
  
  \label{fig:lift} 
\end{figure*}

Commercial buildings also often contain elevators, which are much less commonplace in residential buildings. \figref{fig:lift} shows the power consumption of an elevator in the academic block of our data set. From \figref{fig:lift}~(a), it can be seen that elevators can draw vastly different amounts of power. Furthermore, due to the small journey duration, the power consumption of the elevator is a short time duration transient with no steady states. The variable power draw of the elevator may depend upon a variety of factors such as distance moved and weight carried, among others~\cite{lifts_2, lifts_1}. Additionally, it can be seen from the histogram of elevator power consumption (\figref{fig:lift}~(b)), that most of the time the elevator consumes close to 400~W which is the base consumption when the elevator is not actively used. \figref{fig:lift}~(c) shows the energy consumption of the elevator by the hour of day, averaged over the duration of our deployment. We can observe that the energy consumption in non-working hours is roughly 50\% of the peak energy consumption (which occurs during working hours). Based on this observation, we believe that there is significant value in optimising elevator energy consumption, especially during non working hours.

\textbf{Insights:} Commercial buildings often house systems such as HVAC and elevators which can draw variable amounts of power. The presence of such variable power loads can further complicate disaggregation, especially for NILM approaches which rely on the identification of steady states. Furthermore, we saw that loads such as elevators have a significant energy consumption even when they are not actively used due to their base power demand. Since this baseload power is always on, we believe 
%that it would be non-trivial to design unsupervised NILM methods to train its model. 
that most low frequency methods would fail to disaggregate this portion of energy.
As such, existing unsupervised NILM techniques have only been applied to appliances which exhibit all of their states during routine operation, and cannot be applied to always on loads which never change state.

\begin{figure}[!htb]
  \centering
  \includegraphics[scale=1]{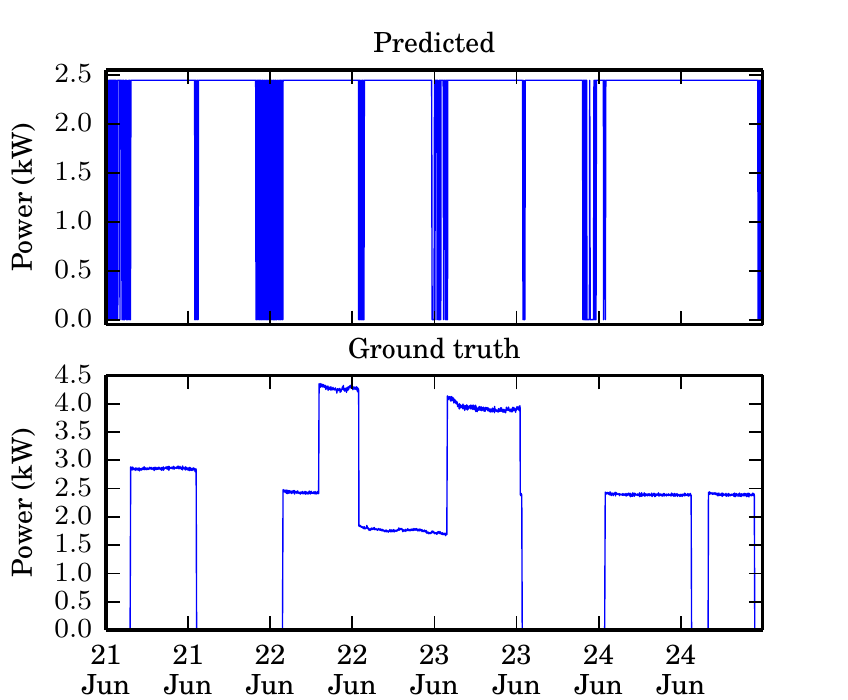} 
  \caption{Predicted power with ground truth for ground floor AHU. The thick bars in the predicted power are due to the frequent on/off switching predicted by CO, which is due to the fact that CO does not model the relationship between different samples.}
  
  \label{fig:disagg_1} 
\end{figure}

\section{Evaluation of disaggregation}\label{evaluation}
Having discussed the traits of NILM which are specific to commercial buildings, we now provide a preliminary evaluation on our data set using benchmark residential NILM algorithms provided by NILMTK~\cite{nilmtk}. We believe that ours is the first evaluation that applies these NILM methods, proposed for residential settings, to a commercial dataset with reasonable sampling time (30 seconds). Firstly, we try to disaggregate AHU loads in the academic block considering the academic block's total power as the input. We use an implementation of the combinatorial optimisation (CO) disaggregation algorithm provided by NILMTK to train the AHUs as two state (on and off) loads. We selected data from the month of June for our experiments and divided it equally into two halves for training and testing. Further, we downsampled the data to 1 minute resolution to synchronise the ground truth and aggregate time series. We use F-score and normalised error in assigned power (NEP) as our accuracy metrics. We follow the standard definition for NEP as defined in NILMTK: the sum of the differences between the assigned power and actual power for each appliance in each time slice, normalised by the appliance's total energy consumption. \tabref{tab:ahu_from_building} shows the disaggregation performance across these AHUs. We observe that the F-score is relatively high suggesting that CO is able to detect the operation of the AHU in most cases. However, \figref{fig:disagg_1} shows an example of the predicted and ground truth power for the ground floor AHU in which CO is unable to identify the AHU operation. This is due to the AHU exhibiting states which were not present in the training set, since the power demand varies with the temperature set point, and thus the finite state CO model is insufficient to accurately disaggregate the AHU power.
%This is due to the variable power demand of the AHU, as discussed in \secref{variability}, wherein the finite state CO model is insufficient to accurately disaggregate the AHU power.

\begin{figure}[!t]
  \centering
  \includegraphics[scale=1]{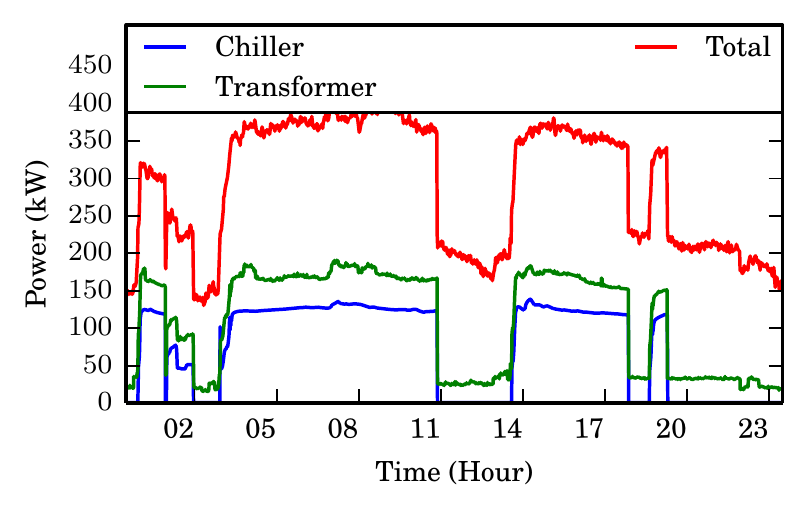} 
  \caption{Chiller dominates the total campus power consumption}
  \label{fig:where} 
\end{figure}

Having discussed disaggregation performance by disaggregating academic block total power into different AHUs, we now see if this performance can be improved with increased sub-metering, as has been shown previously for residential buildings~\cite{batra_2013}. First, we compare the AHU disaggregation performance when disaggregation is performed at the floor level instead of the building level. \tabref{tab:disagg_different_metering} shows that the F-score of the fifth floor AHU improves from 0.7 to 0.99 and the NEP significantly improves from 1.0 to 0.12. Thus, by traversing an additional level in the metering hierarchy, we get near perfect disaggregation accuracy for the AHU. However, we observed that the disaggregation performance remains unchanged for the chiller irrespective of whether we perform the disaggregation at the campus total level or the transformer level. This is likely due to the fact that the chiller already dominates the campus total power consumption and therefore traversing an additional level to the transformer adds little value as shown in \figref{fig:where}. Our observations can also be explained from an information theoretic perspective. \tabref{tab:disagg_different_metering} shows the entropy of the power signal measured at different locations. The disaggregation performance for the AHU improves significantly due to the large decrease in entropy moving from the building level to the floor level. On the other hand, there is insignificant change in the entropy moving from campus total to transformer, which is reflected in the disaggregation performance remaining the same. We believe that such information theoretic approaches can provide useful insights into the minimal set of sensors required to disaggregate a set of desired loads in a given building.

It is also interesting to investigate the disaggregation of the elevator's power consumption from the building total. In this case, we applied the event-based disaggregation approach proposed by Hart~\cite{hart_1992} due to the relatively low elevator power consumption in comparison with the building's HVAC load.
%We failed to obtain a successful disaggregation due to the fact that different streams were not time synchronised, i.e. building total may have readings on the 5th and the 35th second and the elevator may have on the 20th and the 50th second. 
However, since the elevator power data were not time synchronised with the building level power data, it was not possible to evaluate the accuracy of disaggregation at this data resolution. Therefore, we downsampled the data to 1 minute intervals before applying the disaggregation algorithm.
Such downsampling has little effect on the disaggregation performance of CO for loads such as AHUs due to their long state durations. However, this downsampling completely obscures the elevator's power demand. Thus, we believe that greater sampling rates are needed to apply any disaggregation approaches to loads such as elevators which exhibit a short term transient power draw.

\tabcolsep=0.2cm
\begin{table}[!t]
\begin{center}
\begin{tabular}{lcc} \hline
\textbf{Load} &   \textbf{F-score} & \textbf{NEP} \\ \hline
AHU 0 &  0.74 &  0.74 \\
AHU 1 &  0.62 &  1.42 \\
AHU 2 &  0.68 &  1.00 \\
AHU 5 &  0.70 &  1.00 \\ \hline
\end{tabular}
\end{center}
\caption{Disaggregation performance of different AHU loads in the academic block}
\label{tab:ahu_from_building}
\end{table}

%\begin{table}[!htb]
%\begin{center}
%\begin{tabular}{lcccc} \hline
%\textbf{Mains} &Academic block &  & Fifth floor\\
%\textbf{Load} &   \textbf{F-score} & \textbf{NEP} &   \textbf{F-score} & \textbf{NEP}\\ 
%AHU 5 &  0.70 &  1.00 & 0.99 & 0.12\\ \hline
%\textbf{Mains}             &   Campus total  &  & Transformer 2\\ 
%\textbf{Load} &   \textbf{F-score} & \textbf{NEP} &   \textbf{F-score} & \textbf{NEP}\\ 
%Chiller &  1 &  0.93 & 1 & 0.93\\ \hline
%\end{tabular}
%\end{center}
%\caption{Disaggregation performance at different metering levels}
%\label{tab:disagg_different_metering}
%\end{table}

\tabcolsep=0.1cm
\begin{table}[!t]
\begin{center}
\begin{tabular}{ccccc} 
\hline
\textbf{Location} & \textbf{Entropy} &\textbf{Load} & \textbf{F-score} & \textbf{NEP} \\
% & \textbf{at location}& &  &  \\

\hline
Academic block & 15.3 & AHU 5 & 0.70 & 1.00 \\
5th floor & 12.2 & AHU 5 & 0.99 & 0.12 \\
Campus total & 17.5 & Chiller &  1 &  0.93 \\
Transformer 2 & 17.1 & Chiller & 1 &  0.93 \\
\hline
\end{tabular}
\end{center}
\caption{Disaggregation performance at different metering levels. We used NPEET toolbox~\cite{ver2000non} for entropy calculations.}
\label{tab:disagg_different_metering}
\end{table}

%For the purpose of illustration of the information added by sub-metering, we propose a naive algorithm for disaggregating lift power consumption. Previously, in section .. we saw that lifts can be considered as loads having a fixed base power consumption and short term variable power transient when they are used. Hart et al. ~\cite{hart} in their seminal work on NILM had proposed using edge detection and matching algorithms to label the power consumption of appliances. To that basic algorithm, we add the intuituion which is based on our obsevration that the lift transient is of a very short duration. Thus, our algorithm seeks to find positive and negative edge pairs of similar power consumption occuring in a short time duration.

\section{Conclusions and Future work}\label{conclusion}

The majority of previous NILM research has focused on residential buildings. However, commercial buildings often consume several orders of magnitude more energy than residential buildings, and thus the return on investment for NILM is expected to be greater. Previous research in the context of residential buildings have made several assumptions: (i) appliance events happen one-at-a-time, (ii) presence of steady states, (iii) appliance usage temporal patterns and (iv) correlations amongst several loads. In order to determine the validity of these assumptions in commercial settings, we studied these assumptions from our campus deployment and contrasted them with existing residential data sets. We found that there is a significant difference in the amount of energy consumed on weekend and weekdays in our campus. Although such temporal dependencies also exist in residential setting, they are much less profound. NILM algorithms for commercial settings can exploit these temporal dependencies to enhance disaggregation accuracy. Commercial buildings usually house much larger number of loads in comparison to residential buildings and thus exhibit an order of magnitude more events than the residential buildings. This is also due to the fact that commercial buildings often house systems such as VFD based HVAC systems which can have a continuously varying power demand. Both of these factors 
%- large number of events and variable power consumption can make disaggregation harder in the commercial settings. 
increase the complexity of energy disaggregation in commercial settings.
Thus, the common assumption of steady state power consumption and one-at-a-time events may not be valid in commercial settings. While correlations amongst appliances can be seen in residential buildings, the effect is much more profound in commercial buildings when various HVAC subsystems turn on and off at the same time. We also discussed that due to these loads all turning on together, the transient generated may significantly increase the peak power consumption of the building and thus cause the company to incur demand charges.

We also performed a preliminary evaluation of NILM on our data set where we tried to disaggregate AHUs from the building total and floor total respectively. The naive CO approach we used fails to model the continuously varying power demand of the AHU. However, the disaggregation performance improves significantly when we perform disaggregation at the floor level. We have also released a subset of our data for public use.

In the future we would like to enhance the NILM-metadata schema~\cite{NILM_metadata} which is used in conjunction with NILMTK~\cite{nilmtk} to support complex metering hierarchies usually present in commercial buildings. We had briefly discussed information theoretic approaches as an indicator for diaggregation performance in our paper. We would like to study these in more detail to be able to come up with models for optimal sensing. Given that our smart meters are capable of collecting data at a higher resolution of 1 Hz, we would like to explore disaggregating loads such as elevators using event based approaches. 

Further, we would also like to exploit existing work in context discovery in smart commercial spaces to increase the disaggregation accuracy~\cite{thakur, softgreen}. Also, it is likely that existing approaches working with low frequency data may not be able to disaggregate low power consuming systems such as lighting. We would like to explore the possibility of using previous work on EMI sensing~\cite{emisense, electrisense} in addition to our low frequency energy disaggregation to get more detailed energy disaggregation. Additionally, we aim to make an end to end system using existing systems such as BAS~\cite{bas} or SensorAct~\cite{sensoract}, which can collect data, disaggregate using NILMTK and perform control action or provide detailed feedback for energy optimisation.
%We also aim to repeat our current experiments on different kinds of commercial buildings such as manufacturing units and verify the validity of our findings. 

\scriptsize
\bibliographystyle{abbrv}
\bibliography{reference}  % sigproc.bib is the name of the Bibliography in this case

\balancecolumns

\end{document}